\documentclass[prl,aps,twocolumn, floatfix, superscriptaddress,showpacs]{revtex4}
\usepackage{amsmath,bm,graphicx}
\usepackage{amssymb}
\usepackage{amsfonts}
\begin{document}

\title{Cooperative rectification in confined Brownian ratchets}

\author{Paolo Malgaretti}
\email[Corresponding Author : ]{paolomalgaretti@ffn.ub.es }
\affiliation{Department de Fisica Fonamental, Universitat de Barcelona, Spain}
\author{Ignacio Pagonabarraga}
\affiliation{Department de Fisica Fonamental, Universitat de Barcelona, Spain}
\author{J. Miguel Rub\'{\i}}
\affiliation{Department de Fisica Fonamental, Universitat de Barcelona, Spain}
\date{\today}

\begin{abstract}
We analyze the rectified motion of a Brownian particle in a confined environment. We show the emergence of  strong cooperativity between the inherent rectification of the ratchet mechanism and the entropic bias of the fluctuations caused by spatial  confinement. Net particle transport may develop even in situations where separately the ratchet  and the geometric restrictions do not give rise to  particle motion.
The combined rectification effects can lead to bidirectional transport depending on particle size, resulting in a new route for segregation. The reported mechanism  can be used to control transport in mesostructures and nanodevices in which particles move in a reduced space.
\end{abstract}


\maketitle

Brownian motors, identified in a variety of conditions ranging from biological~\cite{Guerin2010} to synthetic systems~\cite{Hanggi2009}, extract work from thermal fluctuations in out of equilibrium conditions.  In particular, Brownian ratchets  rectify thermal fluctuation due to their interaction with a periodic asymmetric potential (ratchet) in a non-equilibrium environment. They constitute a reference class of Brownian motors and  have been widely used to understand how molecular~\cite{Guerin2010,Astumian} as well artificial~\cite{Allison2002, Zhu2004, Linke1999} motors operate.  Geometrical constraints provide an alternative means to rectify thermal fluctuations due to the confinement they impose,  reducing the system capability to explore space. Modulations in the available explored region lead to  gradients in the system  effective free energy,  inducing a local bias in its diffusion  that can promote  a macroscopic net velocity for asymmetric channel profiles~\cite{rubi2010} or due to applied alternating fields~\cite{Wambaugh}. 
Geometric barriers constitue  a common feature at small scales; they are found in a variety of systems, including  molecular transport in zeolites~\cite{zeolites}, ionic channels~\cite{ion_channel_faraudo}, or in microfluidic devices~\cite{Bezrukov,Fujita}, where their shape explains, for example, the magnitude of the rectifying electric signal observed experimentally~\cite{hanggi_2001}. 

Brownian motors usually  operate in  spatially restricted environments where  thermal rectification is affected by the geometrical constraints. Understanding such interplay is then relevant in a  variety of experimental situations  ranging from micrometric systems, like microfluidic devices \cite{Bezrukov,Fujita} or colloids moving in optical tweezer arrays~\cite{grier}, to nanometric conditions, as realized with molecular motors~\cite{Astumian,Guerin2010}, down  to the atomic scale where optical trapping allows to manipulate cold atoms~\cite{Dion}.

In this Letter, we will show that the interplay between a Brownian ratchet and the geometrical constraints  it experiences strongly affects the out of equilibrium dynamics of small particles  and promote cooperative particle transport even when neither the Brownian ratchet nor the geometrical confinement  rectify on their own. We will clarify that such  net current   results from the cooperative rectification of  thermal fluctuations by  the Brownian ratchet and the geometrical constraint and that it has a significant signal-to-noise ratio that allows to experimentally test such motion on length scales comparable to a  few ratchet periods.  
We will also show how the interplay between Brownian and confinement rectification  can lead to a significant enhancement in the rectified   mean velocity  and, depending on particle size, to velocity reversal providing a novel mechanism for particle segregation at the micro- and nano-scale. 

\begin{figure}
 \includegraphics[scale = 0.29,angle=0]{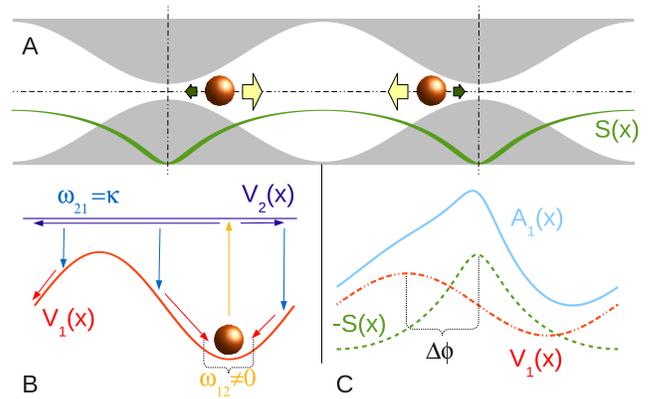}
 \caption{Brownian ratchet and entropic barriers. A: local biased diffusion due to confinement, described in terms of an effective entropic potential, $S(x)$. B: Two-state model for a Brownian ratchet: In $state 1$ particles slide along the potential $V_1$ and jump with  rate $\omega_{1,2}$ to $state 2$ where they diffuse until they jump back with rate $\omega_{2,1}$. The jump rates operate in different  regions along the potential period, breaking detailed balance. C: A Brownian motor moving in a confined environment will be sensitive to the free energy $A_1$ (solid) generated by ratchet potential  $V_1$ (dot-dashed) and entropic potential generated by the channel shown in panel A (dashed).
\label{fig:energetic_entropic_barrier}
}
\end{figure}

In order to identify the generic features underlying the cooperative rectification provided by spatial confinement and Brownian ratcheting, we will analyze the dynamics  of a single particle  of radius $R$ moving in a channel with variable  half-width $h(x)$, where $x$  stands for the position along the channel longitudinal axis, as sketched in Fig.~\ref{fig:energetic_entropic_barrier}.  Rather than analyzing explicitly  the  diffusion of a  particle in  such a channel under the action of a ratchet potential with  the appropriate boundary conditions,  it proves insightful to  take advantage of the asymmetry of the channel geometry and describe the particle dynamics in terms of its  displacement along the channel longitudinal axis and include the channel  boundary as an entropic potential  $\tilde{S}(x)  = k_BTS(x)= k_BT \ln 2(h(x)-R)/R$~\cite{Reguera2001} the particle is subject to ($k_B$ stands for the Boltzmann constant and $T$ is the absolute temperature). 
 Such an approximation, known as Fick-Jacobs~\cite{zwanzig,Reguera2006,Kalinay2008}, is  exact for a  uniform channel
 while its regime of validity for gently  varying confining  geometries has been  explicitly elucidated~\cite{Burada2007}. This approach has been shown  very fruitful  to understand particle transport  in a variety of confined systems~\cite{Reguera2001,ion_channel_faraudo}. 

To address the impact of entropic restrictions on the Brownian motor motion, we will analyze both  a flashing ratchet~\cite{Reimann2002} , a generic model for  the rectified motion of colloidal particles, and the two-state ratchet~\cite{Julicher1997}, modeling the rectified motion of molecular motors along biofilaments which accounts effectively for the   mechanochemical coupling characteristic of molecular motors~\cite{Qian2004}. 

A colloidal particle subject to a periodic external potential, $V_1(x)$ expressed in units of the thermal energy $k_BT$,
 behaves as a flashing ratchet when the random force breaks detailed balance~\cite{grier}. This can be simply achieved with a Gaussian  white noise with a second moment amplitude $g(x)=\sqrt{D(x)+Q\left(\partial_{x}V_{1}(x)\right)^{2}}$~\cite{Reimann2002}, where $Q$ quantifies Brownian rectification.  The particle density, $p(x)$, reads 
\[\frac{\partial p}{\partial t}  = \frac{\partial}{\partial x}\left\{\frac{1}{2}\left[D(x)+Q\left(\frac{\partial}{\partial x}V_1\right)^2\right]\frac{\partial p}{\partial x}+ D(x) p\frac{\partial A_1}{\partial x}\right\},\]
where  the dimensionless free energy $A_{1}(x)=V_{1}(x)- S(x)$, includes the  entropic potential the particle is subject to due to the change in the number of available states as the channel width varies. The channel corrugation  induces a position-dependent  diffusion coefficient, $D(x)$ which depends on the channel section, $h(x)$~\cite{Reguera2001}.

The two-state ratchet model constitutes a standard, simple framework to describe  molecular motor motion. As sketched in Fig.~\ref{fig:energetic_entropic_barrier}, a Brownian particle jumps between two states,  $i=1,2$, which determine under which potential, $V_{i=1,2}$, it displaces~\cite{Julicher1997}.  A choice of the jumping rates $\omega_{12,21}$ that breaks detailed balance, jointly with an asymmetric potential $V_{1}(x)$, determines the average molecular motor velocity $v_0\neq0$.  The conformational changes of the molecular motors  introduce an additional  scale which will compete with  rectification and geometrical confinement.  Infinitely-processive molecular motors remain always attached to the filament along which they displace and are affected by the geometrical restrictions only while displacing along the filament; accordingly, we choose  channel-independent binding rate $\omega_{21,p}(x) = k_{21}$. On the contrary, highly non-processive molecular motors detach frequently  from the  biofilament  and diffuse away, leading to a channel-driven binding rate $\omega_{21,np}(x)=k_{21}/h(x)$. 
The interaction between the molecular motor and the  biofilament is chosen for specificity as 
\begin{equation}
V_1(x)=V_0\left[\sin\left(2\pi x\right)+\lambda\sin\left(4\pi x\right)\right],\partial_{x} V_2=0
\label{energetic_potential}
\end{equation}
in units of $k_BT$,  where the position along the filament, $x$, is expressed in units of the ratchet period, $L$.  $\lambda$ determines the asymmetry of the ratchet potential, $V_1$, while  $V_2$ ensures free diffusion in state $i=2$. Motors jump to the free state only in a region of width $\delta$ around the  minima of  $V_1$,  with rate $\omega_{12}  =   k_{12}$. Accordingly, the motor densities, $p_1,p_2$ along the channel follow~\cite{Julicher1997}
\begin{eqnarray}
\partial_{t}p_{1}(x)+\partial_{x}J_{1} & = & -\omega_{12}(x)p_{1}(x)+\omega_{21}(x)p_{2}(x) \nonumber\\
\partial_{t}p_{2}(x)+\partial_{x}J_{2} & = & \omega_{12}(x)p_{1}(x)-\omega_{21}(x)p_{2}(x) \nonumber
\end{eqnarray}
where $J_{1,2}(x)  =  -D(x)\big(\partial_{x}p_{1,2}(x)+p_{1,2}(x)\partial_{x}A_{1,2}(x)\big)$ stands for the current densities in each of the two states in which motor displaces. Depending on the motor internal state, two dimensionless free energies,  $A_{1,2}(x)=V_{1,2}(x)-S(x)$, account for the interplay between the biofilament interaction and the channel constraints.
\begin{figure}
 \includegraphics[scale = 0.41]{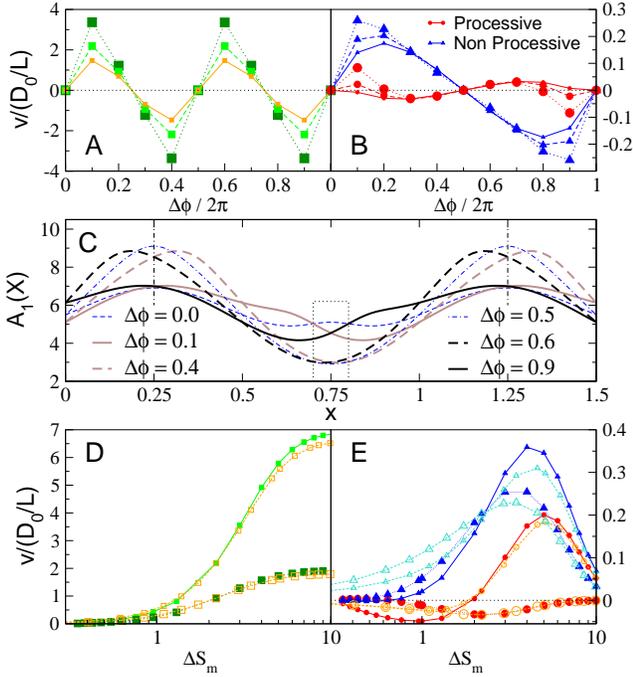}
 \caption{Rectification of a Brownian motor in a symmetric ratchet and symmetric corrugated channel. Panel A,B: velocity, in units of $D_{0}/L$, where $D_{0}=k_BT/6\pi\eta R$ and $L$  stands for the potential period, of a flashing ratchet (squares) or a processive (circles), non-processive (triangles) motor moving according to the two-state model as a function of the phase-shift $\Delta \phi$  for different values of the parameter $\beta/\gamma=0.7,0.8,0.9$ (the larger the symbol   size, the larger $\beta$) and fixed $\gamma$ being the energetic potential $\Delta V_1 = 4.0$ and $\Delta Q=10$ for the flashing ratchet. Panel C: free energy profile for different $\Delta \phi$; the dotted box is the region where  $\omega_{12}\neq0$, the vertical dashed-dotted lines mark the position of the maxima of $A_1(x)$ in the absence of rectification. Panel D-E: rectified velocity as a function of the entropic barrier height upon variation of $R$ (filled points) and $\beta$ (empty points) for flashing ratchet (squares) or a processive (circles), non-processive (triangles) motor moving according to the two-state model. Bigger points stand for bigger $\Delta \phi$ being $\Delta \phi=0.1,0.2$ for the non-processive two-state model and flashing ratchet, $\Delta \phi=0.1,0.3$ for the processive two-state model.}
\label{fig:simm_simm_phase_shift}
\end{figure}

The mean particle velocity is computed from the numerical solution of the   Fokker-Planck equations stated above getting  for the flashing ratchets $v_0=LZ^{-1}\left(1-e^{\phi(L)}\right)$ and for the two-state model  $v_0=(J_1+J_2)L$, where $Z=\int_{0}^{L}dx\frac{e^{-\phi(x)}}{g(x)}\int_{x}^{x+L}\frac{e^{\phi(y)}}{g(y)}dy$ and $\phi=\int_{0}^{x}\frac{D(y)V'_1(y)}{g(y)^{2}}dy$. 

To analyze the interplay between the ratchet potential and the entropic constraints, we will discuss a channel with the same periodicity as the ratchet. In particular, we consider that the channel half-width obeys
\begin{equation}
h(x)  = \gamma+\beta\left[\sin\left(2\pi x+\Delta\phi\right)+\Lambda\sin\left(4\pi x+\Delta\phi\right)\right]
\label{channel}
\end{equation}
where $\Delta\phi$ accounts for the phase difference between the ratchet  and the entropic potentials while $\Lambda$ quantifies  the channel asymmetry. In turn,  $\gamma$ and $\beta$, together with the particle radius $R$, control  the entropic barrier height because the maximum change in entropy reads  $\Delta S_m=\ln \frac{h(x)_{max}-R}{h(x)_{min}-R} $, 
where $(h(x)-R)$ the effective half-section  a tracer of radius $R$ is sensitive to. 

Although when both  the channel and the ratchet are  symmetric, $\lambda = \Lambda = 0$,  none of them can rectify on its own, the interplay of these two mechanisms leads to a net current whose magnitude, controlled by the phase shift  $\Delta \phi$, depends on the relative position of the ratchet along the channel,  as   shown in Fig.~\ref{fig:simm_simm_phase_shift}. Cooperative rectification emerges from the entropy-driven asymmetry of the hopping rates from a minimum of the effective  free energy, $A_1(x)$, to its closest minimum. Confronting panels A,B with panel C in Fig.~\ref{fig:simm_simm_phase_shift} shows that  rectification develops  only when the minimum of the free energy $A_1(x)$ is not equidistant from the free energy maxima, {\sl i.e.} for $\Delta \phi \neq n\pi$ ($n=0,1$). While for the flashing ratchet the net current is always in the direction of the shortest path the particle has to diffuse to overcome the free energy barrier, the intrinsic mechanism of two-state model leads to more involved dynamics. The interplay between internal  motor reorganization and the  geometrically varying environment  can lead to qualitatively new scenarios, such  as the reversal in the direction of motion of processive motors, as clearly shown in panel E of Fig.~\ref{fig:simm_simm_phase_shift}. This scenario, sensitive to the effective cross section felt by the motor as quantified by the parameter $\beta$, cannot be obtained with flashing ratchets, or with alternative ratcheting mechanisms which do not incorporate the internal morphological changes of the displacing particle. 

Even though the free energy, $A_1(x)$, depends explicitly on the channel section, $h(x)$, panels D,E of Fig.~\ref{fig:simm_simm_phase_shift} show that the  dependence  of rectified motion of a confined particle on the channel asymmetry, $\beta$, (open points) and particle size, $R$ (filled points),  are essentially captured when expressing the  rectified velocity in terms of the maximum entropy difference (or channel aperture), $\Delta S_m=\ln \frac{\gamma-R+\beta}{\gamma-R-\beta}$. Only for very small values of $\Delta S_m$, approaching the limit of validity of  the Fick-Jacobs approximation that prescribes a faster thermalization along the radial direction respect to the longitudinal convection~\footnote{ A Brownian ratchet performs an effective motion along the channel when  the  time, $\tau_d = [\partial_xh(x)\Delta x]^2/D_0$, needed to explore the channel section is smaller than the characteristic drifting time,  {\sl i.e.}  $\partial_xh(x)^2 \ll \frac{k_BT}{L\eta\dot{x}}$}, the details of the channel shape and particle size affect quantitatively the net particle velocity.  Therefore, the relative aperture of the channel, quantified by $\Delta S_m$, captures the main dependence of the particle velocity emerging from cooperative rectification and determines, for all the ratchet models explored,   the optimal regime for cooperative  rectified motion.  
\begin{figure}
 \includegraphics[scale = 0.41]{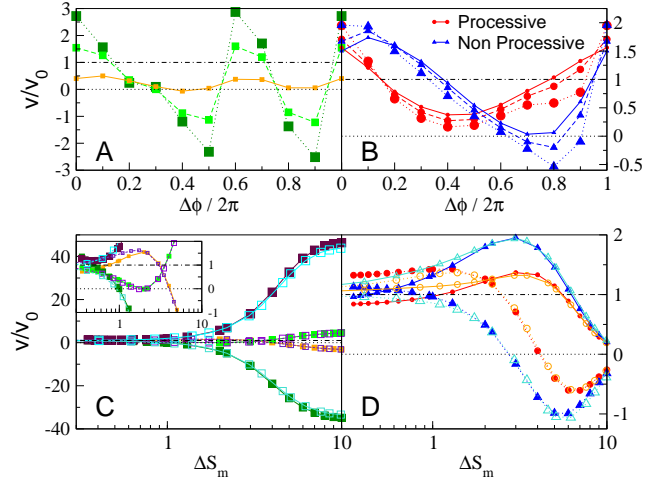}
 \caption{Rectification of a Brownian motor in an asymmetric ratchet and symmetric corrugated channel. Panel A-B: velocity is normalized by $v_{0}$, that is the intrinsic velocity of a particle with the same radius $R$ under the action of the ratchet in the case of a flat channel with the same volume, symbols and parameters values as in figure \ref{fig:simm_simm_phase_shift}. C-D: 
velocity as a function of $\beta$ (open points) or $R$ (fileed points); points as in fgure\ref{fig:simm_simm_phase_shift}. $\Delta \phi=0.2,0.3,0.5,0.6$ for the flashing ratchet and $\Delta \phi=0.1,0.8,0.9$ for the two-state model, the larger the symbols the  bigger $\Delta \phi$.}
\label{resume2}
\end{figure}

The rectified velocities displayed in  Fig.\ref{fig:simm_simm_phase_shift}  can be of order $D_0/L$ for a ratchet potential of magnitude 
 $4k_BT$ {\sl i.e.}  in experimentally achievable regimes. In fact, for optically-driven colloids  ratchet potential amplitudes  one order of magnitude larger than the  thermal energy can be achieved tuning the laser beam intensity while the height of the energy barrier molecular motors are subject to due to ATP hydrolysis can be as much as  $\sim 10k_BT$~\cite{Howard}. 

For asymmetric ratchets, $\lambda \neq 0$, active transport leads to net rectification even for a symmetric, $ \Lambda=0$, corrugated channel. The particle sensitivity to  the channel shape leads now to  strong mean particle velocity enhancement.  Moreover,  panels A,B of Fig.~\ref{resume2} show that   confinement allows  particles  to move against the underlying ratchet rectification. The inset of panel C and panel D of Fig.~\ref{resume2} emphasize the non-monotonic behavior of the velocity with respect to the entropy barrier leading to maxima of the velocity enhancement and inversion. Hence, $\Delta S_m$ allows to identify, generically, the optimal regime for rectified transport. Such a behavior is  stronger for a flashing ratchet  than for molecular motors, leading to velocities up to 40 times larger than  their unconfined counterparts.
Since $\Delta S_m$ depends on the  tracer size (filled points), panels C,D of Fig.~\ref{resume2} indicate  a novel mechanism for particle segregation based on particle size; depending on the phase-shift, $\Delta \phi$, bigger particles can move faster or slower than smaller ones and, by fine tuning the parameters, they can be also trapped or even move in opposite direction.  Finally, if particles displace in the presence of  both an asymmetric  ratchet potential  and channel corrugation ($\lambda\neq0,\Lambda\neq0$)  the strong enhancement in the rectified particle velocity is  kept and both the  inversion in  the direction of motion and  the mechanism for particle size segregation are generically observed for appropriate values of the ratchet parameters. 

In conclusion, we have studied the cooperative rectification between geometrical constraints and Brownian ratchets as a new mechanism for active transport in confined environments and have shown that their interplay  profoundly affects the net motion of small particles.  We have clarified the physical origin of such a rectification, which may take place even when separately neither  entropic nor ratcheting  can lead to net particle motion, elucidating the role played by the biased diffusion generated by the geometrical constraint. The rectified velocity can be detectable in experimentally feasible situations and can be strongly enhanced by increasing the amplitude of the ratchet barrier in the two-state model or the multiplicative noise parameter $Q$ in the flashing ratchet.
In the presence of ratchet rectification,  the entropic constraints modulates the  velocity in a particle-size dependent manner that  leads to regimes of maximum velocity enhancement and velocity reversal, providing a new mechanism for particle segregation in confined environments. Although   cooperative rectification relies on the  phase difference between the ratchet potential and the corrugated channel, the particle velocity vanishes only when in  registry. According to  Fig.~\ref{fig:simm_simm_phase_shift}, a probability distribution of phase shifts, $p(\Delta \phi)$, will still lead to rectification whenever $\sqrt{\langle\Delta \phi^2\rangle}\ll \pi$, with a magnitude which will depend on $\langle\Delta \phi\rangle$. For larger phase shift distributions, $\sqrt{\langle\Delta \phi^2\rangle} \sim \pi$, cooperative rectification will survive only for asymmetric ratchets, with  $\lambda \neq 0$, and will be generically observed when both the entropic and ratchet potentials are asymmetric,  $\lambda \neq 0,\Lambda \ne 0$.   The cooperative  mechanism described is robust, and can be exploited to control active transport of particles in ionic channels~\cite{ion_channel_faraudo} or nuclear pores, confined colloids~\cite{grier} or  nanobead transport  in microfluidic devices. 

We acknowledge  the Direcci\'on General de Investigaci\'on (Spain) and DURSI project for financial support
under projects  FIS\ 2008-04386 and 2009SGR-634, respectively. J.M. Rub\'{\i} acknowledges financial support from {\sl Generalitat de Catalunya }under program {\sl Icrea Academia}

\bibliography{letter_miguel}

\end{document}